\documentclass[prd,preprintnumbers,showpacs,12pt]{revtex4}
\usepackage[active]{srcltx}
\usepackage{epsfig}
\usepackage{graphicx}
\usepackage{bm}
\usepackage{amsmath}
\usepackage{amssymb}

\newcommand{\be}{\begin{equation}}

\newcommand{\ee}{\end{equation}}
\newcommand{\bea}{\begin{eqnarray}}
\newcommand{\eea}{\end{eqnarray}}

\def\f{\frac}
\newcommand{\ep}{\epsilon} 
      
\newcommand{\tgam}{\tilde\gamma}   
\newcommand{\dx}{{\dot x}}
\newcommand{\bx}{{\bf x}}
\newcommand{\bp}{{\bf p}}

\newcommand{\bb}{{\bf b}}
\newcommand{\br}{{\bf r}}
\newcommand{\bxi}{{\bm\xi}}

\newcommand{\nn}{\nonumber}
\newcommand{\de}{\partial}

  \def\slash#1{\setbox0=\hbox{$#1$}#1\hskip-\wd0\dimen0=5pt\advance
        \dimen0 by-\ht0\advance\dimen0 by\dp0\lower0.5\dimen0\hbox
          to\wd0{\hss\sl/\/\hss}}

\usepackage[latin1]{inputenc}
\usepackage{color}
\begin{document}         

\title{Two interacting conformal Carroll particles}. 

\author{Roberto Casalbuoni$^ {a)}$}\email{casalbuoni@fi.infn.it}
\author{Daniele Dominici$^ {a)}$}\email{dominici@fi.infn.it}
\author{Joaquim Gomis$^ {b)}$} \email{joaquim.gomis@ub.edu}
\affiliation{$^ {a)}$Department of Physics and Astronomy, University of Florence and
INFN, 50019 Florence,  Italy}
\affiliation{$^ {b)}$Departament de F\'isica Qu\`antica i Astrof\'isica \\
and Institut de Ci\`encies del Cosmos (ICCUB), Universitat de Barcelona,  Martí i Franquès 1, 08028 Barcelona,
Spain}

\begin{abstract} In this note we analyse two different models of two interacting 
conformal Carroll particles that can be obtained as the 
Carrollian limit of two relativistic conformal particles. The first model describes particles with zero velocity and exhibits 
infinite dimensional symmetries which are  reminiscent of the BMS symmetries.
 A second model of interaction of Carrollian particles is proposed, where
the particles have non zero velocity and therefore, as a consequence of the limit $c\to 0$,  are tachyons.  Infinite dimensional symmetries are present  also in this model.

\end{abstract}

\maketitle   


\section{Introduction}
Carroll symmetry, which was introduced by \cite{LevyLeblond:1965,gupta1966analogue} as the limit of the
Poincaré symmetry where
 the velocity of light is going to zero, $c\to 0$,   has recently received a lot of attention,  mainly in connection with its relation with the BMS group \cite{Bondi:1962px,Sachs:1962zza}.
Some applications of 
Carrollian physics 
 allow,
  for example, to understand the symmetries of null hypersurfaces, such as black-hole horizons \cite{Donnay:2019jiz}, boundaries of asymptotically flat spacetimes \cite{Duval:2014uva}, Carroll fluids \cite{Ciambelli:2018wre}. For an introductory review of certain aspects of Non-Lorentzian 
theories, see \cite{Bergshoeff:2022eog}.

In particular, in this paper we are interested to have a dynamical understanding of the relation between the
conformal Carroll structures \cite{Duval:2014uva} and the symmetries of Carrollian particles.

A (massive) non-conformal Carroll particle
was introduced  as Carroll limit of relativistic massive particle
\cite{Bergshoeff:2014jla}; it can also be obtained from a coadjoint orbit of the Carroll group \cite{Duval:2014uoa}. The (massless) Carroll particle was also considered in \cite{Bergshoeff:2014jla}.
A model of two non-conformal interacting particles was also considered where the individual particles move but the center of mass does not move.

 The symmetry of massive and massless  free Carroll particles is infinite dimensional. In the massless case it contains the finite conformal Carroll symmetry introduced in \cite{Bagchi:2016bcd}, see also \cite{Chen:2021xkw}.
 In the case of the two non-conformal particles the symmetry algebra is finite dimensional \cite{Bergshoeff:2014jla}.
 

The causal structure of the Carroll space allows space-like intervals and therefore the existence of Carroll tachyons. In \cite{deBoer:2021jej} a model  is constructed where a tachyonic Carroll particle with zero energy moves.

In this paper  we propose two models for two conformal Carroll particles as Carrollian limits
of two conformal relativistic particles introduced in  \cite{Casalbuoni:2014ofa}. 

The first one  describes  particles which do not move and exhibits an  infinite dimensional symmetry algebra.
 Any  free conformal Carroll particle has vanishing energy, therefore two free Carrollian particles have total energy equal to zero. The nice feature of this model is that the system can have a  total energy different from zero and still be Carroll conformal invariant. Clearly the energy, which turns out to be constant, depends on the interaction.
We also study some possible connections with the infinite dimensional conformal symmetry of
\cite{Duval:2014uva} and the BMS symmetry. 

In the second  model the
particles move,  their individual  energy is zero and consequently they are Carrollian tachyons. Also in this case the symmetry turns out to be  infinite dimensional.

\section{Conformal  Carroll particle}
\label{carrollparticle}

The canonical form of the action of a massless relativistic particle  is given by
(we assume the signature of the metric $(+,-,-,-)$)
\be\label{ads}
S=\int d\tau [ -p\cdot \dot x-\frac {e}{2}p^2]\,,
\ee
where the dot denotes the differentiation with respect to an invariant parameter $\tau$.
 The action is also invariant under the conformal group that contains dilatations and special conformal transformations (SCT). The infinitesimal transformations are given by the ordinary Poincaré
transformations, plus dilatations
\be
\delta x^\mu=\ep_D x^\mu,\,\,\,\,\delta p^\mu=-\ep_D p^\mu,\,\,\,\delta e=2 e \ep_D.
\label{dil}
\ee
and  SCT:
\be
\delta x^\mu=b^\mu x^2 - 2(b\cdot x)x^\mu,\,\,\,\,\delta p^\mu=-2 b\cdot p x^\mu+2 b\cdot x p^\mu+2 p\cdot x b^\mu,
\label{sct1}
\ee
\be
\delta e=-4e  b\cdot x.
\label{sct2}
\ee

We next consider the  Carroll limit which  is given by

\be
x^0= \frac{t}{\omega}, \quad p_0=\omega E, 
\ee
with $\omega\to\infty$. It is understood that, before taking the limit,  we rescale the einbein variable like
\bea
e\to\frac {e}{\omega^2}.
\eea
{Note that we use a dimensionless parameter $\omega$  instead of the velocity of light. Indeed, if one considers the Carrollian counterpart of the non-relativistic limit of a string,  one needs to use a dimensionless parameter \cite{Gomis:2000bd,Gomis:2005pg,Barducci:2018wuj}.}
The contractions we consider in this work correspond to the  limit $c\to 0$  of a world probed by particles. There are more general possible contractions  that correspond to the non-Lorentzian limits of extended objects such as strings and branes.

 In the Carroll limit  the action  (\ref{ads}) becomes
\begin{eqnarray}\label{0}
S_{C} &=&  \int d\tau\, \big[ - E \dot{t}
+ \bp\cdot\dot{\bx}  -\frac{e}{2} E^2\big]\,.
\end{eqnarray}
From the symplectic form of the canonical action we deduce the following Poisson brackets
\be
\{E,\ t\} = 1, \quad \{e, \  \pi\} = 1,\quad \{x^i,p^j\} = \delta^{ij}\, \quad i=1,2,3.
\ee

This action can also be obtained by the method of non-linear realisations \cite{Bergshoeff:2014jla}
and coadjoint orbit method  
\cite{Duval:2014uoa}. 
Since the action (\ref{ads}) is relativistic conformal invariant, the action 
\eqref{0} is also invariant under the Carroll conformal transformations.
 We have
\bea\label{Ctransf'}
&&\delta t={\bm\beta}\cdot {\bf x}+a_t\,,\hskip 2truecm
\delta x^i= \epsilon^{ijk}\theta^j x^k+a^i\,,
\nn\\
&& \delta p^i= \epsilon^{ijk}\theta^j p^k +{\beta^i} E\,, \hskip 2.9truecm
\delta E=0\,,
\eea

\be
\delta t=\ep_D t,\,\,\,\,\delta\bx=\ep_D \bx,\,\,\,\delta E=-\ep_D E,
\delta \bp=-\ep_D \bp,\,\,\,\delta e=2 \ep_D e\,,
\label{carrolldil0}
\ee
\be
\delta t=-b^0 \bx^2+ 2 \bb\cdot \bx\, t,\,\,\,\,\delta\bx=-\bb\, \bx^2+ 2(\bb\cdot \bx) \,\bx\,,
\label{carrollSCT0}
\ee
\be
\delta E=-2 \bb\cdot \bx E,\,\,\,
\delta \bp=-2(b^0E -\bb\cdot \bp)\bx-2 \bb\cdot\bx\, \bp+ 2(Et-\bp\cdot \bx)\bb\, ,
\label{carrollSCT2}
\ee
\be
\delta e=4\bb\cdot \bx e\, ,
\label{carrollSCT3}
\ee
where $\beta^i, a_t, a^i,\epsilon_D, 
b^0,b^i$ are the infinitesimal parameters
associated to Carrollian boost, space-time
translations, dilatation and special Carrollian conformal transformations.

The phase space realization of the conformal Carroll generators  is the following
\be
H={ E}\,,\hskip 1truecm {\bf P} = \bp\hskip 1truecm {\bf G} = E\bx\,,\hskip 1truecm {\bf J} = \bx \times \bp\,,
\ee
\be
D=-Et+\bp\cdot \bx,\quad K^0=-E \bx^2,\quad {\bf K}=2 D \bx -\bx^2 {\bp}\,.
\ee
The corresponding transformations are obtained as  $\delta =\{ G,\,\,\}$.
 
The  conformal Carroll algebra  in phase space in four dimensions is  given by (see \cite{Bagchi:2016bcd}, and also \cite{Chen:2021xkw}) 

\begin{eqnarray}\label{Calgebra}
&&\{E, P^i\} =\{E,G^i\}= \{E,J^i\}=\{E, K^0 \} = 0\,,\{E, D\} = -E\,,\{E, K^i\} = -2 G^i\,,\nn\\
&&\{P^i,P^j\}=0\,, \{P^i,G^j\} = -\delta^{ij}E\,,\{P^i,J^j \} = \epsilon^{ijk} P^k\,,\nn\\
&&\{P^i, D\} = -P^i\,,\{P^i, K^0 \} = 2G^i\,,\{P^i, K^j\} = 2 \epsilon^{ijk} J^k-2D\delta^{ij}\,\,,\nn\\
&&\{G^i,G^j\}=0\,, \{G^i,J^j, \}= \epsilon^{ijk}G^{k}\,,\{G^i, D \} = 0\,,\{G^i, K^0\} =0\,,\{G^i, K^j \} =\delta^{ij} K^0\,,\nn\\
&&\{J^{i}, J^{j}\}= \epsilon^{ijk} J^{k}\,,\{J^i, D\} = 0\,,\{J^i, K^0\} = 0\,,\{J^i, K^j \} = \epsilon^{ijk}K^k\,,\nn\\
&&\{D, K^0\} = -K^0\,,\{D, K^i\} = -K^i\,,\nn\\
&&\{K^i, K^0\} = 0\,,\{K^i,K^j\}=0\,.
\end{eqnarray}

 The conformal Carroll algebra can be obtained from the relativistic conformal 
 algebra with generators $M^{\mu\nu}, P^\mu,
 K^\mu, D$ associated to Lorentz, space translations, SCT and dilatations
 \bea
\left[\tilde M^{\mu\nu},\tilde M^{\rho\sigma}\right]&=&\eta^{\nu\rho}\;\tilde M^{\mu\sigma}+...,
\nn\\
\left[\tilde P^\rho,\tilde K^\sigma\right]&=&2\,(\eta^{\rho\sigma}\;\tilde D\;+\;\tilde M^{\rho\sigma}\;),
\nn\\
\left[\tilde P^{\mu},\tilde M^{\rho\sigma}\right]&=&\eta^{{\mu}[\rho}\;\tilde P^{\sigma]},
\nn\\
\left[\tilde K^{\mu},\tilde M^{\rho\sigma}\right]&=&\eta^{{\mu}[\rho}\;\tilde K^{\sigma]},
\nn\\
\left[\tilde P^\rho,\tilde D\right]&=&-\;\tilde P^\rho,
\nn\\
\left[\tilde K^{\rho},\tilde D\right]&=&\tilde K^{\rho},
\nn\\
\left[\tilde P^{\mu},\tilde P^\nu\right]&=&\left[\tilde K^{\mu},\tilde K^\nu\right]=\left[\tilde D,\tilde M^{\mu\nu}\right]=0\\
 \eea
   by the following contractions
 \be
  E=\f 1\omega\, \tilde P^0, \, G^i=\f 1 \omega \,\tilde M^{i0},\,
 K^0=\f 1\omega \tilde K^0\,
 \ee
 and the identification
 \be
 J^i=\f 1 2  \epsilon^{ijk} \tilde M^{jk}
 \ee

\subsection{Infinite-dimensional Symmetries}

Now we want to analyse  which is
 the most general point transformation of the Carroll particle action (\ref{ads}) following the steps of 
\cite{Bergshoeff:2014jla}.

In order to do that we want to write the Carollian Killing equations. Let us first consider the equations of motion  deduced from (\ref{0})

\be
\dot t=-eE,\quad \dot{\bf{x}}=0,\quad \dot e =\lambda(\tau), \quad \dot E=0, \quad \dot{\bf{p}}=0\,, \quad\dot\pi=-1/2E^2\,,
\label{eq:16}
\ee
where $\lambda (\tau)$ is an arbitrary function and $\pi$ is the momentum associated to the einbein
variable  which is constrained by  $\pi=0$.

Consider the following generator of the canonical transformations 
\be
G=E\xi^0(\bx,t)-\bp\cdot \bxi (\bx,t)+\gamma(\bx,t)\pi,
\ee
with  $\xi^0(\bx,t)\,,\xi^i(\bx,t)$ and $\gamma(\bx,t)$ arbitrary functions of the space-time coordinates.
 The transformations generated by this generator  are given by
\bea
\delta t&=\{G,t\}=&\xi^0(\bx,t),\quad \delta x^i=\{G,x^i\}=\xi^i(\bx,t), \quad \quad \delta e=\{e,G\}=\gamma(\bx,t)\,,
\nn\\[.1truecm]
 \delta  E&=\{G,E\}=&  \partial_t \xi^0(\bx,t) E+\partial_t \xi^i(\bx,t)p_i+\partial_t \gamma(\bx,t)\pi\,,
\nn\\[.1truecm]
 \delta p^i&=\{G,p^i\}=&\partial_i\xi^0(\bx,t) E +\partial_i\xi^j(\bx,t) p_j+\partial_i\gamma(\bx,t)\pi\,.
\eea

These transformations are  symmetries of the free Carroll particle, provided that $G$ is a constant of
motion, i.e., $d G/d\tau=0$.This leads to the following restriction, after use of Eq.~(\ref{eq:16}) and $\pi$ that is a primary constraint:
\bea
&0&=E \left [\dot t\partial_t\xi^0(\bx,t)+\dot{x_j}\partial^j\xi^0(\bx,t)\right ]+
p_i  \left [\dot t\partial_t\xi^i(\bx,t)+\dot{x_j}\partial^j\xi^i(\bx,t)\right ]+\dot\pi \gamma(\bx,t)\nonumber\\[.1truecm]
&& =-eE^2\partial_t\xi^0(\bx,t)-eEp_i\partial_t\xi^i(\bx,t)-\frac{1}{2}\gamma(\bx,t)E^2\,.
\eea
From this equation we deduce the following Killing equations corresponding to the free Carroll particle \cite{Bergshoeff:2014jla}:
\be
\partial_t \xi^0(t,\bx)=-\frac{\gamma}{2e}\,,\quad \partial_t\xi^i(t,\bx)=0
\ee
for arbitrary parameter $\gamma(\bx,t)$.
This leads to the following  generator of the conformal  Killing transformations
\be
G=E\xi^0(t,\bx)+p_i\xi^i(\bx)-2\pi e \partial_t \xi^0(t,\bx)\,,
\ee
which generates an infinite dimensional symmetry, which will be called ${\cal G}_1$.   These transformations include Carroll conformal  transformations. However they are more general. The dependence of $\xi^0(t,\bx)$ on the parameter $t$ is arbitrary while in BMS is linear \cite{Duval:2014uva}. This arbitrary dependence on $t$ is reminiscent of the Newman-Unti group, which contains BMS which is isomorphic to the infinite extension of the Carroll conformal group \cite{Duval:2014uva}.

%

%
%
%

\section{Two conformal  relativistic
particles}
In this section we  briefly recall the most relevant features of a  model of two interacting conformal particles proposed in \cite{Casalbuoni:2014ofa}. The canonical action is formulated, in terms of the coordinates $x_1,x_2$, the momenta $p_1,p_2$, the einbeins $e_1,e_2$
and the associated momenta $\pi_1, \pi_2$ that are primary
constraints, by 
\bea\label{tworelcan}
S
&=& \int d\tau \left(-p_1{\dx}_1-p_2\cdot {\dot x}_2+\pi_1 {\dot e_1}+
\pi_2 {\dot e_2}
 -H\right)\,,\label{eq:3.6}\eea
 where the  Hamiltonian $H$ is given by
\bea
H&=&
-e_1\phi_1-e_2\phi_2
-\pi_1 \mu_1-\pi_2 \mu_2\,,
\eea
 $\mu_1,\mu_2$ are arbitrary functions of $\tau$ and the mass-shells constraints are
\bea\label{massshelltworelativistic}
&&\phi_1= \frac 12\left(p_1^2-\frac{\alpha^2}4\sqrt{\frac{e_2}{e_1}}\frac 1 {r^2}\right),\nn\\
&&\phi_2=\frac 12\left(p_2^2-\frac{\alpha^2}4\sqrt{\frac{e_1}{e_2}}\frac 1 {r^2}\right)\,
\eea
where $r=x_1-x_2$  it is relative space-time coordinate.
 Note that when the interaction, which is regulated by the parameter $\alpha$, vanishes the model describes  two massless free relativistic particles.
 Notice also  that the model is conformal invariant with  effective masses different from zero, 
 a property  that is not possible for a single conformal particle.

%
The equations of motion are given by
\be
\dot x_a^\mu=\{x^\mu_a,H\}=e_a p_a^\mu,\,\,\,a=1,2,
\ee
\be
\dot p_1^\mu=\{p^\mu_1,H\}=\f {\alpha^2}{4}\sqrt{e_1e_2}\{p^\mu_1,\f {1}{r^2}\}=-\sqrt{e_1e_2}\f {\alpha^2}{2r^2}r^\mu=-\dot p_2^\mu,
\ee
\be
\dot e_a=\{e_a,H\}=\mu_a\,\,\,a=1,2,
\ee
\be
\dot \pi_a=\{\pi_a,H\}=\phi_a,\,\,\,a=1,2\,.
\ee
The stability of the primary constraints $\pi_a$ implies the secondary constraints $\phi_a=0$.
 Note the appearance of the einbeins
in the  constraints $\phi_a$.
\subsection{ Killing equations and symmetries}
 Like in the free particle case we would like to find the most general point transformations of the model. We want to write the corresponding Killing equations for the two particle model. In order to do that we write the most general generator

\be
G=\sum_{a=1}^2\left [ \xi_{a\mu}(x_1,x_2)p_a^\mu+\gamma_a \pi_a\right ].\label{eq:5.1}\ee
As shown in \cite{Casalbuoni:2014ofa} the requirement
$
 {d G}/{d \tau}=0
$
is satisfied provided the following conditions are verified
\be
\xi_a(x_1,x_2)=\xi_a(x_a),\,\,\, a=1,2,
\ee
\be
\f {\de \xi_{a\mu}}{\de x_a^\nu}+\f {\de \xi_{a\nu}}{\de x_a^\mu}-g_{\mu\nu} \tgam_a=0,\,\,\,a=1,2,
\label{xicond1}
\ee
where  
\be
\tgam_a=\f {\gamma_a} {e_a},\,\,\, a=1,2,
\ee
and
\be
(\xi_1-\xi_2)\cdot r=\f {r^2} 8 \sum_{a=1}^2\de_{a\nu}\xi_a^\nu.
\label{xicondtwo}
\ee

From Eq.~(\ref{xicond1}) we also obtain
\be
\tgam_a=\f  1 2  \de_{a\nu}\xi_a^\nu, \,\,\, a=1,2.
\ee

We have verified that the model is invariant under the diagonal subgroup of  the two conformal groups $SO(4,2)_{1,2}$ acting on the two variables $x_1$ and $x_2$ respectively by checking that Eqs.~(\ref{xicond1}) and ~(\ref{xicondtwo}) are satisfied for a generic infinitesimal transformation of the diagonal conformal subgroup, 
\be
\xi_a^\mu=\ep^\mu +\omega^\mu_{a\nu} x^{\nu}_a+\ep_{D} x_a^\mu +b^\mu x_a^2 - 2(b\cdot x_a)x^\mu_a,
  \,\,\, a=1,2.
\ee

We can now compute the transformation properties of the variables under dilatations and SCT
using
\be
\delta x_a^\mu=\{G,x^\mu\}=\xi_a^\mu,\,\,\,\,\delta p_a^\mu=\{G,p_a^\mu\}=-\frac {\de \xi_{a}^\nu}{\de x_{a\mu}}{p_{a\nu}},\,\, a=1,2,
\ee
\be
\delta e_a=\{e_a,G\},\,\, a=1,2.
\ee

{Under dilatations:}

\be
\delta x_a^\mu=\ep_D x_a^\mu,\,\,\,\,\delta p_a^\mu=-\ep_D p_a^\mu,\,\,\,\delta e_a=2 e_a\ep_D,\,\, a=1,2.
\label{dil}
\ee

{Under  special conformal transformations:}
\be
\delta x_a^\mu=b^\mu x_a^2 - 2(b\cdot x_a)x^\mu_a,\,\,\,\,\delta p_a^\mu=-2 b\cdot p_a x_a^\mu+2 b\cdot x_a p_a^\mu+2 p_a\cdot x_a b^\mu,\,\, a=1,2,
\label{sct1}
\ee
\be
\delta e_a=e_a\tgam_a=-e_a 4 b\cdot x_a,\,\, a=1,2.
\label{sct2}
\ee

\section{Two conformal  Carroll
particles}
Here we  consider the Carroll limit of the conformal relativistic canonical action 
(\ref{tworelcan}) by assuming

\be
{p_a}^0=\omega E_a,\,\,\,x_a^0= \f 1 \omega t_a, \,\,\,a=1,2
\ee
From the mass-shell constraints given in Eq.~(\ref{massshelltworelativistic}) we get
\bea
&&\phi_1= \frac 12\left(\omega^2 E_1^2 -{\bf p}_1^2+\frac{\alpha^2}4\sqrt{\frac{e_2}{e_1}}\frac 1 {\f 1 {\omega^2}r_0^2-{\bf r}^2}\right),\nn\\
&&\phi_2= \frac 12\left(\omega^2 E_2^2 -{\bf p}_2^2+\frac{\alpha^2}4\sqrt{\frac{e_1}{e_2}}\frac 1 {\f 1 {\omega^2}r_0^2-{\bf r}^2}\right)\,.
\eea
By defining
\be
{\tilde e_a}=e_a \omega^2,\,\,\, {\tilde \alpha}^2=\f {\alpha^2}{\omega^2},\,\,\,a=1,2,
\ee
we obtain, in the limit $\omega\to\infty$,
\bea
e_1\phi_1&=& 
 \frac  {{\tilde e}_1}2\left(E_1^2 -\frac{{\tilde\alpha}^2} 4\sqrt{\frac{{\tilde e}_2}{{\tilde e}_1}}\frac 1 {{\bf r}^2}\right),
 \quad 
\eea
\be
e_2\phi_2=\frac  {{\tilde e}_2}2\left(E_2^2 -\frac{{\tilde\alpha}^2} 4\sqrt{\frac{{\tilde e}_1}{{\tilde e}_2}}\frac 1 {{\bf r}^2}\right)\,,
\ee
so that  the Carroll canonical Lagrangian is given by
\bea
L&=&
-E_1\dot t_1+{\bf p}_1\cdot {\dot\bx}_1-E_2\dot t_2+{\bf p}_2\cdot {\dot\bx}_2\nn\\
&&+\frac  {{\tilde e}_1}2\left(E_1^2 -\frac{{\tilde\alpha}^2} 4\sqrt{\frac{{\tilde e}_2}{{\tilde e}_1}}\frac 1 {{\bf r}^2}\right)
+\frac  {{\tilde e}_2}2\left(E_2^2 -\frac{{\tilde\alpha}^2} 4\sqrt{\frac{{\tilde e}_1}{{\tilde e}_2}}\frac 1 {{\bf r}^2}\right)\,,
\label{Lcarroll}
\eea
From now on we neglect the tilde, by renaming $\tilde e\to e,\,\,\,\tilde \alpha\to \alpha$.

When $\alpha=0$ the Lagrangian is invariant under two independent Carroll transformations acting on the corresponding particle coordinates. 
When $\alpha\neq 0$, the Lagrangian is invariant under the following diagonal Carroll transformations for each particle
\be
\delta_C x^i_a=\epsilon^i -\Lambda^{ij}x^j_a,\,\,\,\delta_C p^i_a= -\Lambda^{ij}p^j_a+\beta^j E_a\,,
\label{eq:46}
\ee
\be
\delta_C t_a=h+{\bm\beta}\cdot {\bf x}_a,\,\,\,\delta_C E_a=0,\,\,\, \delta e_a=0,\,\,\,a=1,2.
\label{eq:47}
\ee
Lagrangian equations of motion are
\bea
\dot\bp_1&=&\f {\alpha^2} 2\f {\sqrt{e_1e_2}}{\br^4}\br,\nn\\
\dot\bp_2&=&-\f {\alpha^2} 2\f {\sqrt{e_1e_2}}{\br^4}\br,\nn\\
\dot\bx_a&=&0,\,\,\,\dot E_a=0,\,\,\,\dot t_a=e_aE_a,\nn\\
0 &=&\dot\pi_1=\frac {\de L}{\de e_1}=\f 1 2 \left (E_1^2-\f {\alpha^2}{4\br^2}\sqrt{\f {e_2}{e_1}}\right )\nn\\
0&=&\dot\pi_2=\frac {\de L}{\de e_2}=\f 1 2 \left (E_2^2-\f {\alpha^2}{4\br^2}\sqrt{\f {e_1}{e_2}}\right )\,.
\label{Piadot}
\eea
This model describes interacting  
 Carroll
particles that have zero velocity.  Since the distance among the particles is constant, it is only fixed by the initial conditions.

\subsection{Eliminating  the einbeins}
The einbeins are non-dynamical variables and can be eliminated through their equations of motion \cite{Pons:2009ch}. 
In fact from the equations of motion of  $e_1$ we obtain
\be
\sqrt{\f {e_1}{e_2}}=\f {\alpha^2}{4E_1^2\br^2}
\ee
and by squaring
\be
 {e_1}={e_2}\left (\f {\alpha^2}{4E_1^2\br^2}\right )^2\,.
\ee
Then the canonical Carroll Lagrangian can be rewritten as
\be
L=-E_a\dot t_a+\sum_{a=1,2}\bp_a\cdot \dot \bx_a +\f {e_2} 2 \left ( E_2^2-(\f {\alpha^2}{4\br^2})^2 \f 1 {E_1^2}\right )\,.
\label{eq:133}
\ee
The equation of motion  for $E_2$ gives
\be
\dot t_2=e_2 E_2
\ee
and substituting in Eq.~(\ref{eq:133}) we obtain
\be
L=-E_1\dot t_1-\f 1 2 \f {\dot t_2^2}{e_2}+\sum_{a=1,2}\bp_a\cdot \dot \bx_a -\f {e_2} 2 (\f {\alpha^2}{4\br^2})^2 \f 1 {E_1^2}\,.
\label{eq:135}
\ee
By using the equation for $e_2$
\be
e_2=\pm \f{4\br^2}{\alpha^2}\left (E_1^2t_2^2\right )^{1/2}\,.
\ee
and substituting in Eq.~(\ref{eq:135}), the Lagrangian can be rewritten as
\be
L=-E_1\dot t_1+\sum_{a=1,2}\bp_a\cdot \dot \bx_a \mp   \f {\alpha^2}{4\br^2} \left (\f {\dot t_2^2}{E_1^2}\right )^{1/2}\,,
\label{eq:138}
\ee
For the equation of motion of  $E_1$ we get
\be
E_1=\pm \f {\alpha}{2}(\dot t_2^2)^{1/4}\f 1 {({\br^2\dot t_1})^{1/2}}\,.
\ee

By substituting $E_1$ in Eq.~\ref{eq:138} the final Lagrangian is (with a suitable choice of $\pm$ sign)
\bea
L&=&\sum_{a=1,2}\bp_a\cdot \dot \bx_a-\alpha \left [\f {{\dot t}_1^2{\dot t}_2^2 }{\br^4}\right ]^{1/4}\nn\\
&=&\sum_{a=1,2}\bp_a\cdot \dot \bx_a-\f  \alpha { |\br |}\left [ {\dot t}_1{\dot t}_2\right ]^{1/2}\,.
\label{eq:131}
\eea
We assume ${\dot t}_1{\dot t}_2>0$. 

 Since the Lagrangian is 
homogeneous of degree one in the velocities,
it is invariant under diffeomorphisms and gives rise to  the primary constraint
\be
E_1E_2-\f {\alpha^2}{4\br^2}=0\,.
\label{eq:62}
\ee

 In the case of two conformal free particles their individual  energies are zero. Instead in this model, when the interaction is turned on ($\alpha\neq 0$),  the energies are different from zero and constant. The particles are interacting, indeed, as shown in Eq.~(\ref{eq:62}), the energy of each particle depend on the energy and the position of the other one.
 In conclusion the model is Carroll conformal  even if the two particles have non-vanishing  energies. 

Since  the energies $E_1$ and $E_2$ are constants of motion, the meaning of the Eq.~(\ref{eq:62}) is that each particle stays inside its own light cone, corresponding to the lines with constant values of $\bx_1$ and $\bx_2$ (the velocities are zero, see eqs. (\ref{Piadot})) and of  $\br$. Introducing the total and  the relative energies of the two particles:
\be
E_T=E_1+E_2,~~~E_r=E_1-E_2\,,
\ee
we get
\be
E_T^2=E_r^2+\f {\alpha^2}{\br^2}\,.
\ee
This equation shows that the minimum of the total energy is for $E_1=E_2=0$ and for $\br\to\infty$.

The  wave equation  associated to the Eq.~(\ref{eq:62}) is
\be
[\partial_{t_1}\partial_{t_2}+\f {\alpha^2}{4\br^2}]\Phi(t_1,t_2,\vec x_1,\vec  x_2)=0
\ee


\subsection{Invariance under dilatations and special conformal transformations}

The Lagrangian (\ref{Lcarroll}) is invariant also under the dilatations and special  conformal transformations.
Under dilatations we have:
\be
\delta t_a=\ep_D t_a,\,\,\,\,\delta\bx_a=\ep_D \bx_a,\,\,\,\delta E_a=-\ep_D E_a,
\delta \bp_a=-\ep_D \bp_a,\,\,\,\delta e_a=2 \ep_D e_a,\,\,\, a=1,2.
\label{carrolldil}
\ee

Under SCT:
\be
\delta t_a=-b^0 \bx_a^2+ 2 \bb\cdot \bx_a\, t_a,\,\,\,\,\delta\bx_a=-\bb\, \bx_a^2+ 2\bb\cdot \bx_a \,\bx_a,\,\,\, a=1,2.
\label{carrollSCT}
\ee

\be
\delta E_a=-2 \bb\cdot \bx_a E_a,\,\,\,
\delta \bp_a=-2(b^0E_a -\bb\cdot \bp_a)\bx_a-2 \bb\cdot\bx_a\, \bp_a+ 2(E_at_a-\bp_a\cdot \bx_a)\bb,\,\, 
\label{carrollSCT2}
\ee
\be
\delta e_a=4\bb\cdot \bx_a e_a,\,\, ,a=1,2.
\label{carrollSCT3}
\ee
These SCT are obtained from the covariant ones given in Eqs.~(\ref{sct1}) and (\ref{sct2}), considering
\be
x_a^0=\f 1 \omega t_a,\,\,\,p^0_a=\omega E_a,\,\,\,b^0=\f {b^0}{\omega},\,\,\, a=1,2.
\ee

\subsection{Infinite dimensional symmetries}
\label{subsC}
Like in the previous sections, here we would like to find the most general point symmetries of the model.


Let  consider 
the generic Killing vector
\be
G=\sum_{a=1,2}\left [\xi^0_a (t_1,t_2,\bx_1,\bx_2)E_a-\bxi_a(t_1,t_2,\bx_1,\bx_2)\cdot \bp_a
+\gamma_a(t_1,t_2,\bx_1,\bx_2)\pi_a\right ]\,.
\ee
By considering the $\tau$ derivative we obtain
\bea
\f {d G}{d \tau}&=&
\sum_{a,c=1,2}\Big [\f {\partial \xi^0_a(t_1,t_2,\bx_1,\bx_2)}{\partial t_c}  \dot t_c E_a
-
 \f {\partial \xi_a^j(t_1,t_2,\bx_1,\bx_2)}{\partial t_c}\dot t_c p_a^j
-\bxi_a(t_1,t_2,\bx_1,\bx_2)\cdot \dot\bp_a\nn\\
&&+\gamma_a (t_1,t_2,\bx_1,\bx_2)\dot\pi_a\Big ]\,,
\eea
where we have used the equations of motion
\be
\dot E_a=0,\,\,\,\dot\bx_a=0\quad 
\ee
and the primary constraints $\pi_a=0$.
Using
Eqs.~(\ref{Piadot}) we get
\bea
\f {d G}{d \tau}&=&\sum_{a,c=1,2}\Big[\f {\partial \xi^0_a(t_1,t_2,\bx_1,\bx_2)}{\partial t_c}  e_c E_c E_a-
 \f {\partial \xi_a^j(t_1,t_2,\bx_1,\bx_2)}{\partial t_c}e_c E_c p_a^j+ \f 1 2 \gamma_a (t_1,t_2,\bx_1,\bx_2)E^2_a\Big ]\nn\\
 &&
 -\f {\alpha^2} {2\br^4} \sqrt{e_1e_2}(\bxi_1-\bxi_2)\cdot \br
-\f {\alpha^2}{8\br^2}\left [\gamma_1(t_1,t_2,\bx_1,\bx_2) \sqrt{\f {e_2}{e_1}}+\gamma_2(t_1,t_2,\bx_1,\bx_2) \sqrt{\f {e_1}{e_2}}\right ]\,.
\label{Gdot}
\eea
Therefore the Killing condition ${d G}/{d \tau}=0$ implies
\be
\f {\partial \xi_a^j(t_1,t_2,\bx_1,\bx_2)}{\partial t_c}=0,\,\,\,a,c=1,2,
\ee
or    $
\xi_a^j=\xi_a^j(\bx_1,\bx_2)
$.
We see that $\delta \bx_a\equiv \bxi_a$ do no depend
on the times $t_1,t_2$.

By defining 
\be
\gamma_a=e_a \tgam_a
\ee
from ${d G}/{d \tau}=0$ we must have also
\be
\f {\partial \xi^0_a(t_1,t_2,\bx_1,\bx_2)}{\partial t_c}=-\f 1 2\delta_{ac}\tgam_a,,\,\,\,a,b,c=1,2,
\label{eq:157}
\ee
and
\be
\f {\alpha^2} {\br^2}(\bxi_1-\bxi_2)\cdot \br=\f  {\alpha^2} 2 \f {\delta {\br^2}}{\br^2}=- \f {\alpha^2} 4 (\tgam_1+\tgam_2),\,\,\,a=1,2.
\ee
In the case $\alpha =0$, that is when we turn off the interaction, we see that  we have two sets of independent Killing equations, one set for each particle. Therefore the infinite group of symmetry transformations is ${\cal G}_1\times{ \cal G}_1$. When the interaction is on and $\alpha\not=0$, we get:
\be
\f 1 {\br^2}(\bxi_1-\bxi_2)\cdot \br=\f 1 2 \f {\delta {\br^2}}{\br^2}=- \f 1 4 (\tgam_1+\tgam_2),\,\,\,a=1,2.
\label{eq:159}
\ee
From Eq.~(\ref{eq:157}) we have
\be
\frac {\de \xi^0_1}{\de t_2}=0=\frac {\de \xi^0_2}{\de t_1}\,.
\ee
Furthermore from Eq.~(\ref{eq:159}) we note that since $\bxi_a$ are only functions of $\bx_1,\bx_2$,
$\tilde\gamma_a$ depend only on $\bx_1,\bx_2$. 
In conclusion we obtain
\be
\xi^0_a\equiv \delta t_a=-\f 1 2 \tilde\gamma_a(\bx_1,\bx_2)t_a+h_a(\bx_1,\bx_2),
\quad \delta e_a =\gamma_a={e_a}\tilde\gamma_a(\bx_1,\bx_2)\,,
\ee
with the extra condition given by Eq.~(\ref{eq:159}).

Summarising the Killing generator is given by
\be
G=\sum_{a=1,2}\left [\big(-\f  1 2\tilde\gamma_a(\bx_1,\bx_2)t_a+h_a(\bx_1,\bx_2)\big) E_a-\bxi_a(\bx_1,\bx_2)\cdot \bp_a
+{e_a}\tilde\gamma_a(\bx_1,\bx_2)\pi_a\right ]\,,
\ee
where $h_a,\bxi_a,\tilde\gamma_a$ are arbitrary functions of $\bx_1,\bx_2$ that must satisfy the condition (\ref{eq:157}).
Therefore the original ten arbitrary functions reduce to nine independent arbitrary functions of $\bx_1,\bx_2$.


Therefore the model of two conformal Carroll particles analysed in this note  has  an infinite dimensional group of symmetries that we will call ${\cal G}_2$. Notice that ${\cal G}_2\subset {\cal G}_1\times{\cal G}_1$. The main difference between these two groups is   the transformation laws of   the times $t_a$ that in the free case are arbitrary functions of the times and the  coordinates, while in the interacting case the Eq.~ (\ref{eq:157}) restricts the time dependence of the transformations of $\xi^0_a$ to be linear in $t_a$.

It is convenient to make a comparison with \cite{Duval:2014uva}.
Let us generalise to two particles  the expression for the generators of the   conformal Carrollian algebra from \cite{Duval:2014uva}. We consider the diagonal subgroup of the two conformal Carroll groups.

For Carroll dilatations and SCT we have 
\be
\xi^i_a=(\epsilon_D+2 \bb\cdot \bx_a)x^i_a-b^i \bx_a^2\,,
\label{eq:170}
\ee
\be
\xi^0_a=(\epsilon_D+ 2 \bb\cdot \bx_a)t_a + T_a(\bx_1,\bx_2)\,.
\label{eq:171}
\ee
We have
\be
\f {\de \xi^0_a}{\de t_a}=\epsilon_D+ 2 \bb\cdot \bx_a
\ee
and
\be
\f 1 2 \left (\f {\de \xi^0_1}{\de t_1}+\f {\de \xi^0_2}{\de t_2}\right )=\epsilon_D+  \bb\cdot \bx_1+ \bb\cdot \bx_2
\ee
\bea
(\bxi_1-\bxi_2)\cdot \br &=&\epsilon_D \br^2+2 (\bb\cdot \bx_1)\bx_1\cdot\br- \bb\cdot \br (\bx_1)^2-2 (\bb\cdot \bx_2)\bx_2\cdot\br+\bb\cdot \br (\bx_2)^2\nn\\
&=&\epsilon_D \br^2+
\br^2(\bb\cdot \bx_1+\bb\cdot \bx_2)\,.
\eea
In conclusion Eq.~(\ref{eq:159}) is satisfied by Eq.~(\ref{eq:170}) and Eq.~(\ref{eq:171}) and the so-called super-translations $T_a(\bx_1,\bx_2) $ in our case are arbitrary functions of both particle positions $\bx_1,\bx_2$.


\section{Two conformal  Carroll particles. A tachyonic model.}
\label{carrolltachyonic}
In this Section, following the approach of \cite{deBoer:2021jej} \cite{Gomis:2022spp}, we develop a model describing two interacting Carrollian tachyons. Let us consider the Lagrangian 
\bea
L &=&
\sum_{a=1,2}\left [-E_a\dot t_a+\bp{\dot\bx}_a -{e_a}\phi_a-\chi_aE_a\right ]\nn\\
&=&
-E_1\dot t_1+{\bf p}_1\cdot {\dot\bx}_1-E_2\dot t_2+{\bf p}_2\cdot {\dot\bx}_2\nn\\
&&-\frac  {{e}_1}2\left(\bp_1^2 -\frac{\alpha^2} 4\sqrt{\frac{e_2}{e_1}}\frac 1 {{\bf r}^2}\right)
-\frac  {{ e}_2}2\left(\bp_2^2 -\frac{{\alpha}^2} 4\sqrt{\frac{{ e}_1}{{e}_2}}\frac 1 {{\bf r}^2}\right)\nn\\
&&-\chi_1 E_1-\chi_2E_2\,.
\label{Lcarroll2}
\eea

Notice the presence of the Lagrangian multipliers $\chi_1, \chi_2$, that implement
that the energies are zero; these terms are necessary in order to have Carroll invariance. An analogous situation appears for single particle in \cite{Gomis:2022spp} where the action is named the magnetic Carroll particle.
In fact 
this Lagrangian is invariant under the Carroll, Eqs.~ (\ref{eq:46}) and (\ref{eq:47}), scale  and SCT,  Eqs.~(\ref{carrolldil}-\ref{carrollSCT3}), by requiring the following transformations for the einbein $\chi_a$
\be
\delta \chi_a=-e_a\bp_a\cdot \bm\beta,\quad a=1,2, \quad {\rm Carroll~~transf.s}\,,
\label{eq:76}
\ee

\be
\delta \chi_a=\epsilon_D \chi_a\quad a=1,2, \quad {\rm scale~~ transf.s}\,,
\label{eq:77}
\ee

\be
\delta \chi_a=2 \left [e_a (b^0\bp_a\cdot \bx_a -t_a \bp_a\cdot \bb)+\chi_a\bb\cdot \bx_a\right ]\quad a=1,2, \quad {\rm SCT~~ transf.s}\,.
\label{eq:78}
\ee
The equations of motion derived from the Lagrangian (\ref{Lcarroll2}) by varying with respect to $t_a,E_a,\bp_a,\bx_a,\chi_a,e_a$ are the following
\be
\quad \dot E_a=0,\quad \dot t_a=-E_a\,,
\label{eq:79}
\ee
\be
e_a\bp_a-\dot\bx_a=0,\quad \dot \bp_1=-\f {\alpha^2}{2 \br^4}\sqrt{e_1e_2} \br=-\dot \bp_2\,,
\label{eq:80}
\ee
\be
E_a=0\,, 
\label{eq:81}
\ee
\be
\phi_a=0\,.
\ee

In this model the particles have non zero velocity and therefore, as a consequence of the limit $c\to 0$,  they are necessarily tachyons. Or equivalently, since $E_a=0$, if we use relativistic mass-shell constraint the two particle invariant squared masses are negative ($E_a^2-\bp_a^2<0$).

\subsection{Eliminating the einbeins}
 Let us first study the equivalence of this model with the $c\to 0 $ limit of the model 
 for two relativistic conformal particles \cite{Casalbuoni:2014ofa} described by the Lagrangian
\be
L=-\alpha \left ( \f {\dot x_1^2\dot x_2^2}{r^4}\right )^{1/4}\,.
\label{lagrtwoconf}
\ee

 In order to do that, we start by eliminating the   non-dynamical variable 
 $e_1$
 in Eq.~(\ref{Lcarroll2}) 
  through its own equation of motion. We obtain  
\be
\sqrt{\f {e_1} {e_2}}= \f {\alpha^2}{4\br^2\bp_1^2}
\ee
and substituting in Eq.~(\ref{Lcarroll2})
\bea
L&=&
-E_1\dot t_1+{\bf p}_1\cdot {\dot\bx}_1-E_2\dot t_2+{\bf p}_2\cdot {\dot\bx}_2\nn\\
&&-\f {e_2}2 \left [\bp_2^2 -(\f {\alpha^2}{4\br^2})^2\f 1 { \bp_1^2}\right ]\nn\\
&&-\chi_1 E_1-\chi_2E_2\,.
\eea

The equations of motion of the non dynamical variables 
 $\bp_2$ are
\be
\bp_2=\f 1 {e_2} \dot \bx_2
\ee
and the Lagrangian becomes
\bea
L&=&
-E_1\dot t_1-E_2\dot t_2-\chi_1 E_1-\chi_2E_2\nn\\
&&+{\bf p}_1\cdot {\dot\bx}_1+\f 1{2 e_2}\dot\bx_2^2+\f {e_2}2  (\f {\alpha^2}{4\br^2})^2\f 1 { \bp_1^2}\,.
\eea

Then we eliminate the non dynamical variable
$e_2$ through its own equation of motion.
(choosing the $-$ sign)
\be
e_2= -\f {4\br^2}{\alpha^2}\sqrt{\dot \bx_2^2\bp_1^2}
\ee
and substituting in $L$
\bea
L&=&-\sum_{a=1,2}(E_a\dot t_a+\chi_aE_a)\nn\\
&&+\bp_1\cdot {\dot\bx}_1-\f {\alpha^2}{4\br^2}\sqrt{\f {\dot \bx_2^2}{\bp_1^2}}\,.
\eea
Finally, taking into account that
\be
\dot \bx_1=-\bp_1 \f {\alpha^2}{4\br^2}\sqrt{\dot\bx_2^2}(\bp_1^2)^{-3/2}
\ee
and
\be
\bp_1^2= \f {\alpha^2}{4\br^2}\sqrt{\f {\dot\bx_2^2}{\dot\bx_1^2}}\,,
\ee
we get the following expression for the Lagrangian
\bea
L&=&-\sum_{a=1,2}(E_a\dot t_a+\chi_aE_a)\nn\\
&&-\alpha \left ( \f {\dot \bx_1^2\dot \bx_2^2}{\br^4}\right )^{1/4}\,.
\label{eq:97}
\eea
By eliminating $E_a$ and $\chi_a$ we obtain
\be
\dot t_a=-\chi_a,\quad E_a=0
\ee
and the following expression for the Lagrangian
\bea
L&=&-\alpha \left ( \f {\dot \bx_1^2\dot \bx_2^2}{\br^4}\right )^{1/4}\,,
\label{eq:97b}
\eea
which coincides with the $c\to 0$ limit of (\ref{lagrtwoconf}).

A primary constraint can be obtained by squaring $\bp_a$
\be
\bp_1^2\bp_2^2-\f {\alpha^2}{16\br^4}=0\,.
\label{eq:100}
\ee

\subsection{ Killing equations and symmetries}
In this Section we derive  the Killing vector by working directly with the Lagrangian given in Eq.~(\ref{Lcarroll2}).
By taking the total derivative with respect to $\tau$ of the Killing vector
\be
G=\sum_{a=1,2 }(\xi_a^0E_a-\bxi_a\cdot \bp_a+\gamma_a\pi_a+\nu_a\pi_a^\chi )\,,
\ee
where
\be
\pi_a^\chi=\f {\de L}{\de \dot \chi_a}\,,
\ee
 we get
\bea
\f {d G}{d \tau}
&=&\sum_{a,b=1,2}\left(\f {\de\xi_a^0}{\de t_b}\dot t_bE_a+\f {\de\xi_a^0}{\de \bx_b}\dot \bx_bE_a-
\f {\de\bxi_a}{\de t_b}\dot t_b\bp_a-\f {\de\xi_a^i}{\de x_b^j}\dot x_b^j p_a^i\right )\nn\\
&&-\sum_{a=1,2}(\bxi_a\cdot \dot\bp_a+\gamma_a\dot \pi_a+\nu_a\dot \pi_a^\chi)\nn\\
&=&\sum_{a,b=1,2}\left(-\f {\de\xi_a^0}{\de t_b}\chi_bE_a+\f {\de\xi_a^0}{\de \bx_b}e_b \bp_bE_a+
\f {\de\bxi_a}{\de t_b}\chi_b\bp_a-\f {\de\xi_a^i}{\de x_b^j}e_bp_b^j p_a^i\right )\nn\\
&&-\sum_{a=1,2}(\bxi_a\cdot \dot\bp_a+\gamma_a \phi_a+\nu_a E_a)\nn\\
&=&\sum_{a,b=1,2}\Big [
 (-\f {\de\xi_a^0}{\de t_b}\chi_b+\f {\de\xi_a^0}{\de \bx_b}e_b \bp_b-\nu_a)E_a-
 \f {\de\bxi_a}{\de t_b}\chi_b\bp_a-\f {\de\xi_a^i}{\de x_b^j}e_bp_b^j p_a^i\nn\\
 &&- \f 1 2\gamma_a \bp^2_a-\f {\alpha^2}{2\br^4}\sqrt{e_1e_2}(\bxi_1-\bxi_2)\cdot \br-\f {\alpha^2}{8r^2}\sqrt{e_1e_2}(\tgam_1+\tgam_2))\Big ]\,,
\eea
where use has been made  of the primary constraints $\pi_a=\pi^\chi_a=0$, of Eqs.(\ref{eq:79}),(\ref{eq:80}) and
of 
\be
\dot\pi_a=-\phi_a\,,
\ee
\be
\dot \pi^\chi_a=-E_a\,.
\ee
The sum over latin indices is understood.

In order to get ${d G}/{d \tau}=0$, we must require
\be
\f {\de\bxi_a}{\de t_b}=0\,,
\ee
so that $\bxi_a$ is a function of only $\bx_1,\bx_2$; we require also the vanishing of the coefficient of $E_a$

\be
\sum_b(-\f {\de\xi_a^0}{\de t_b}\chi_b+\f {\de\xi_a^0}{\de \bx_b}e_b\bp_b )-\nu_a=0\,.
\ee

We also require, assuming $\bxi_a=\bxi_a(\bx_a)$, the vanishing of the coefficient of   $p_a^ip_a^j$
\be
\left ( \f {\de \xi^i_a}{\de x_a ^j}+\f {\de \xi^j_a}{\de x_a ^i}\right )=-\delta^{ij}\tgam_a(\bx_a)\,,
\ee
so that ${d G}/{d \tau}$ implies, when $\alpha\neq 0$
\be
-\f 1 4(\tgam_1+\tgam_2)=\f 1 6 \left ( \f {\de \xi_1^i}{\de x_1 ^i}+\f {\de \xi_2^i}{\de x_2 ^i}\right )=\f {(\bxi_1-\bxi_2)\cdot \br}{\br^2}\,.
\label{eq:107}
\ee
In conclusion the general solution for the Killing vector is
\be
G=\sum_{a=1,2 }\left [\xi_a^0(t_1,t_2, \bx_1,\bx_2)E_a-\bxi_a(\bx_a)\cdot \bp_a
+e_a\tgam_a\pi_a+\nu_a\pi_a^\chi \right ]\,,
\ee
where
\be
\tgam_a=-\f 2 3 \f {\de \xi^i_a}{\de x_a ^i},\quad
\ee
\be
\nu_a=\sum_b (-\f {\de\xi_a^0}{\de t_b}\chi_b+\f {\de\xi_a^0}{\de \bx_b}e_b \bp_b )\,,
\label{nua}
\ee
with $\xi_a^0$ arbitrary function of $t_1,t_2, \bx_1,\bx_2$ and $\bxi_a(\bx_a)$ satisfying Eq.~(\ref{eq:107}).
  As expected from the tachyonic Carroll particle case discussed in Section \ref{carrolltachyonic}, while $\xi^0_a,\bxi_a,\tgam_a$ are arbitrary point functions of $t_a$ and $\bxi_a$, in general $\nu_a$ depend also on the momenta $\bp_a$.  Carroll, scale transformations and conformal transformations satisfy Eq.~(\ref{eq:107}).

We have checked that, by using the expression of $\xi_a^0$ for Carroll, scale and SCT in Eq.~(\ref{nua}), we recover the correct transformations for $\chi_a$ given  in Eqs.~(\ref{eq:76}),(\ref{eq:77}),(\ref{eq:78}), by computing
\be
\delta\chi_a=\{\chi_a, G\}
\ee
and assuming
\be
\{\chi_a,\pi^\chi_b\}=-\delta_{ab}
\ee
Note also that the dependence of $\nu_a$ on $\bx_a, \bp_a$ in principle could give additional contributions to the transformations of $\bp_a, \bx_a$, which however  are vanishing because proportional to the primary constraints $\pi_a^\chi$.

Notice also  that in the case of two non-conformal particles \cite{Bergshoeff:2014jla} the individual particles move with energies different from zero and they have only a diagonal finite conformal symmetry.

\section{Conclusions}

Carroll symmetries have recently received a lot of attention because of several applications in different domains of theoretical physics, from theory of gravity and strings to studies of fractons in condensed matter.

In this paper, after reexamining the case of a single conformal Carroll particle, we derive the conformal Carroll generators, their algebra and the infinite extension of this algebra \cite{Bergshoeff:2014jla}. Then we propose two different models of Carroll interacting particles that provide  dynamical realisations of the Carroll conformal algebra. The models are obtained from the relativistic model of interacting conformal particles proposed in \cite{Casalbuoni:2014ofa} through suitable limits for $c\to 0$.

Concerning the dynamics, the first model describes particles with zero velocity but with non vanishing energy as a consequence of the interaction. Free conformal Carroll particles have zero energy; here, when the interaction is on,  the energy of one particle depends on the energy of the second one and on the particle relative distance which is constant. 

The second model is a tachyonic one: the particles have non zero velocity and therefore, due to the limit $c\to 0$, they are necessarily tachyons. Nevertheless their energy is vanishing.

Both models exhibit, after a complete analysis of the most general point symmetry transformations, infinite dimensional symmetries which include super-translations as in the case of the BMS group \cite{Bondi:1962px,Sachs:1962zza}, the symmetry that arises in asymptotically flat space-time at null infinity. These infinite dimensional algebras  contain the Carroll conformal one  \cite{Duval:2014uva}, which are equivalent to BMS ones.

In the future several directions could be investigated: on the one hand it would be interesting to extend the analysis of the two conformal Carroll models to include
corrections to the next order in  $ c\to 0$ expansion, see for example \cite{Gomis:2022spp},
 on the other hand a deeper analysis of the infinite symmetries could be performed to check similarities with the BMS group.
 
\acknowledgements
We would like to thank interesting discussions
with Carles Batlle, Josep Pons  and  Axel Kleinschmidt for a careful reading of the first version version of this paper.
Also JG acknowledges the warm hospitality at the Max Planck Albert Einstein Institute where this work was completed. 
The work of JG has
been supported in part by MINECO FPA2016-76005-C2-1-P and PID2019-105614GB-C21
and from the State Agency for Research of the Spanish Ministry of Science and Innovation
through the Unit of Excellence Maria de Maeztu 2020-203 award to the Institute of Cosmos
Sciences (CEX2019-000918-M).

\end{document}